\documentclass[12pt]{article}
\usepackage{paper2e}
\usepackage{mydefs2e}

\newcommand{\B}{{\it B}\xspace}

\begin{document}
\begin{titlepage}
\preprint{UMD-PP-97-59\\
{\tt hep-ph/9611387}\\
November, 1996}

\title{Compositeness and Supersymmetry Breaking\\\medskip
in the Observable Sector}

\author{Markus A.~Luty}

\address{Department of Physics\\
University of Maryland\\
College Park, Maryland 20742, USA\\
{\tt mluty@physics.umd.edu}}

\begin{abstract}
We consider models for physics beyond the standard model in which
\susy is broken spontaneously near the weak scale by fields that
are charged under \ew symmetry.
We show that this is possible if some or all of the light quarks and
leptons are composite near the weak scale.
Flavor-changing neutral currents
can be naturally suppressed by a GIM mechanism or by approximate 
flavor symmetries.
\CP and \B violation may be suppressed by accidental symmetries.
We give a general effective field theory analysis of such models, and 
argue that they can be phenomenologically acceptable and lead to 
interesting observable signals in future experiments.
We then construct explicit models based on non-perturbative effects
discovered by Seiberg.
\end{abstract}


\end{titlepage}

\section{Introduction}
\Susy is arguably the most attractive framework for solving the 
gauge hierarchy problem, the problem of why the weak scale 
$M_{W} \sim 100\GeV$ is so much smaller than scales such as
$M_{\rm GUT} \sim 10^{16}\GeV$ and $M_{\rm Planck} \sim 10^{19}\GeV$
that are believed to play a fundamental role in physics.
If \susy is realized in nature, it must be broken, and 
naturalness arguments indicate that the scale of \susy breaking
(measured by masses of the superpartners of observed particles)
cannot be much larger than the weak scale.
A complete theory that uses \susy to solve the hierarchy problem
must therefore explain why the scales of \ew and \susy breaking are 
related.
The traditional approach as been to break \susy in a ``hidden 
sector'' that is uncharged under the \ew gauge group.
The information that \susy is broken is then transmitted to the 
observable fields through a ``messenger sector.''
The most popular choices in the literature for the messenger sector
are gravitational interactions \cite{SUGRA} or weak gauge interactions 
\cite{GaugeMediated,NewGaugeMediated}.
In such models, the fact that the \ew breaking scale is close to the 
\susy breaking scale can be explained by the mechanism of radiative 
symmetry breaking \cite{Radiative}.

In this paper, we will explore the alternative that \susy and \ew
symmetry are spontaneously broken in the observable sector, that is,
by fields that transform under \ew symmetry.
The simplest explanation of the relation between the \susy and \ew 
breaking scales is then that \susy and \ew symmetry are both broken 
by the same dynamics at the same scale.
Such models are usually not considered because general results for
weakly-coupled theories imply that they lead to unacceptably light
superpartner masses \cite{SSum,DimopoulosGeorgi}.
We point out that these difficulties can be avoided if
some of the ordinary quarks are composite (strongly interacting)
near the weak scale.
We perform an effective field theory analysis to show that such 
models are phenomenologically viable, and predict interesting 
observable phenomena in future experiments.
Flavor-changing neutral currents can be naturally suppressed by a
GIM mechanism, or by approximate flavor symmetries.
These models also have the potential to solve the \susc \CP and
\B violation problems through accidental symmetries.

We then turn to model building.
Composite models of quarks based on non-perturbative effects in \susc 
gauge theories have been recently constructed in 
\Refs{NelsonStrassler}.%
\footnote{In these models, there are multiple compositeness scales 
linked to the quark flavor structure, and \susy breaking is not
addressed.}
As emphasized in these papers, it is exciting that the 
recent advances in understanding non-perturbative effects in \susc 
gauge theories allows us to construct composite models without
\emph{ad hoc} dynamical assumptions.
A different class of composite models have also been advocated
recently in \Ref{EffSUSY}.
We construct explicit composite models that break \susy in the
observable sector, in accordance with the general analysis described
above.
While these models are not fully realistic, they do reproduce many of
the gross features of the real world, and they illustrate both the 
attractive features and the difficulties in constructing realistic
models of this type.

\section{Effective Field Theory Analysis}
In this section, we describe the general features we expect from a 
model in which \susy and \ew symmetry are broken at the same scale.
We first show that such models are possible if quarks are
composite.
Our starting point is the theorem of Dimopoulos and Georgi
\cite{DimopoulosGeorgi}, which states that in a general \susc model at
tree level there is always one first-generation squark mass eigenstate
with mass at most $m_{d} \sim 10\MeV$.%
\footnote{This can be viewed as a refinement of the more general 
sum rule of \Ref{SSum}.}
In order to avoid this phenomenological disaster, loop effects must
be important.
If the theory is weakly interacting, loop effects are important 
only if tree-level effects are suppressed.
This is the case in hidden sector models, where superpartner 
splittings in the observable sector vanish at tree level.
This paper will explore the alternative possibility that loop effects
are important because the theory is strongly interacting at the scale 
where \susy is broken.

The existence of the light squark guaranteed by the theorem of 
\Ref{DimopoulosGeorgi} depends only on the validity of 
the tree-level mass formula for the ordinary quarks, so avoiding this 
result requires that some of the light quark fields are strongly 
interacting at a scale $\La$ near the weak scale.
Standard-model interactions are weak at this scale, so this scenario
necessarily involves new strong interactions.
One possibility is that the quarks carry extra quantum numbers
corresponding to a strong gauge interaction that is broken at the scale
$\La$.
Another possibility is that quarks are tightly-bound composite 
particles made of ``preons'' bound by confining interactions.
These physical pictures are actually equivalent (``complimentary'')
in some models \cite{compliment}, and there may well be other
possibilities.
We follow general practice and say that the quarks are ``composite''
at the scale $\La$ if they are strongly interacting at that scale,
independently of the nature of the strong interactions.
Below the scale $\La$, effects of compositeness can be summarized by
an effective lagrangian containing higher-dimension operators
suppressed by powers of $1/\La$.

Bounds on flavor-conserving 4-fermion operators give \cite{PDG}%
\footnote{The interpretation of bounds on flavor-violating
higher-dimension operators depends on the structure of the flavor sector,
and will be addressed below.}
\beq[CompositeBound]
\La \gsim \hbox{few~TeV}.
\eeq
We will assume that this is sufficiently large that the compositeness
effects can be parameterized by higher-dimension operators suppressed
by powers of $1/\La$ in the effective lagrangian at the weak scale.

\subsection{Superpartner Masses}
In the effective theory below the compositeness
scale $\La$, the effective \spot is highly constrained by 
non-perturbative renormalization theorems \cite{NonPertNonRenorm},
but the effective \Kahler potential is believed to contain all terms
that are consistent with the symmetries of the underlying theory.
We now show that if \susy is broken in the effective theory below
the scale $\La$,  then terms in the effective \Kahler potential can 
give rise to superpartner masses for quarks and leptons.
This mechanism was discussed previously in \Ref{EffSUSY}.

For example, if the low-energy theory contains singlet superfields,
there will be terms in the effective \Kahler potential 
of the form
\beq[SquarkKeff]
\de K\rsub{eff} = \frac{c}{\La^{2}} S^{\dagger} S Q^{\dagger} Q,
\eeq
where $S$ is a singlet field.
If $S$ is a composite field, then we expect the coefficient $c \sim 1$.
If $S$ is an elementary field that couples to strongly-coupled
``preons'' $P$ via
superpotential terms of the form
\beq[Scoup]
\de W \sim \la S P P,
\eeq
then $c \sim \la$, which can also be of order 1.%
\footnote{Note that in \Eq{Scoup} $\la S$ appears as a 
field-dependent mass term for the preons, and \Eq{SquarkKeff} can be
viewed as a mass-dependent correction to the effective lagrangian for 
the composite fields analogous to the terms in the effective 
lagrangian for pions that depends on the quark masses.
Just as in the pion effective lagrangian, the coefficient is not 
suppressed by loop factors.}
In order to avoid suppression by small couplings or loop factors,
\Eq{SquarkKeff} must be invariant 
under the larger group of symmetries that results when all 
interactions that are weak at the compositeness scale are turned off.

If the $\th\th$ component of $S$ gets a \vev $F$ (of mass dimension 2),
then \Eq{SquarkKeff} gives rise to a squark mass
\beq[SquarkMass]
m_{\twi{Q}}^{2} \sim \frac{F^{2}}{\La^{2}}.
\eeq
Experimental observations require
$m_{\twi{Q}} \gsim M_{W}$, so that
$F / \La \gsim M_{W}$.
If \susy is broken in the observable sector, naturalness requires
$F \lsim (4\pi M_{W})^{2} \sim (1\TeV)^{2}$,
which gives the bound
$\La \lsim 16\pi^{2} M_{W} \sim 10\TeV$.

We see that \susy must be broken near the compositeness scale, where the 
theory is strongly interacting.
The simplest explanation for this is that \susy is broken by the 
strong composite dynamics, and $F \sim \La^{2}$.
In this case,
\beq
\La \sim \sqrt{F} \lsim 4\pi M_{W} \sim 1\TeV.
\eeq
If $F \sim \La^{2}$, higher-dimension operators involving 
additional powers of $F / \La^{2}$ will not be suppressed, and
the existence of terms such as \Eq{SquarkKeff} simply reflects the 
fact that the squark masses are not protected once \susy is broken.
We will continue to write operators such as \Eq{SquarkKeff} even in 
models where $F \sim \La^{2}$, but it should be kept in mind that this is
only to show that some effect is allowed by all symmetries and
not suppressed by any small parameter.

Note that we have no guarantee \emph{a priori} that the signs of
coefficients such as $c$ in \Eq{SquarkKeff} are such that color and
electomagnetism are unbroken.
In the absence of any known dynamical principle, 
we will simply assume that a phenomenologically acceptable vacuum
is obtained.

Gaugino masses can arise from terms such as
\beq[GauginoKeff]
\de K_{\rm eff} = \frac{g^{2}}{\La^{3}}
S_{1}^{\dagger} S\sub{2} \tr(W W) + \hc,
\eeq
where $S_{1,2}$ are singlet fields.
(We will explain the use of two different singlets below.)
If the $\th\th$ components of $S_{1,2}$ have \vevs of order $F$,
this gives rise to gaugino mass of order
\beq
m_{\twi{g}} \sim \frac{g^{2} F^{2}}{\La^{3}}.
\eeq
Note that the gaugino mass is suppressed by a factor of $g^{2}$ 
compared to squark masses (for $F \sim \La^{2}$).
This suppression is appreciable only for the gaugino corresponding 
to the gauge group $\U1_{Y}$ ($g_{Y}^{2} \sim 0.1$).
It is therefore likely that the Bino is the next-lightest \susc 
particle (the lightest is the gravitino).

There is no loop suppression factor if $S_{1,2}$ are composite, or if
$S_{1,2}$ are elementary fields coupled to strongly-interacting fields
as in \Eq{Scoup}.
A loop suppression factor would give gaugino masses of order
$g^{2} \La / (16\pi^{2}) \sim 5\GeV$, which is certainly too light for 
winos.

As already discussed above, we must make sure that \Eq{GauginoKeff} is 
invariant under all symmetries that exist when the weak couplings are 
turned off.
In particular, the strong gauge dynamics that leads to compositeness
will generally have an anomaly-free 
$\U1_{R}$ symmetry in this limit, and so the combination
$S_{1}^{\dagger} S\sub{2}$ must have $R = -2$.
(This is why we did not use $S^{\dagger} S$ in \Eq{GauginoKeff}.)
One must therefore be careful that $\U1_{R}$ symmetries do not suppress 
the gaugino masses.
This is something that must be checked in each individual model.

\subsection{Flavor-changing Neutral Currents}
In any extension of the standard model, we must consider the
possibility of \FCNCs.
For example, if the squark masses have arbitrary flavor structure,
they will in general give rise
to unacceptably large \FCNCs through loop effects.

One way to suppress \FCNCs is to assume that the theory above the scale 
$\La$ possesses a GIM mechanism \cite{GIM}
similar to the one discussed in the context
of technicolor theories in \Refs{TechniGIM}.
This idea is quite general, and we describe two variants of
the idea below.
When we construct specific models, we will see that approximate flavor
symmetries
(similar to those considered in \Refs{FlavorScalars,FlavorScalarstoo})
can also play a role in suppressing \FCNCs.

In the absence of quark masses, the standard model has a flavor 
symmetry
\beq
G\rsub{flavor} = \SU{3}_{Q} \times \SU{3}_{U} \times \SU{3}_{D},
\eeq
under which the quark fields transform as
\beq
Q \sim (\Yfund, 1, 1),
\quad
\bar{U} \sim (1, \bar{\Yfund}, 1),
\quad
\bar{D} \sim (1, 1, \bar{\Yfund}).
\eeq
This flavor symmetry must be broken by couplings with the same
``spurion'' transformation under $G\rsub{flavor}$ as quark
masses:
\beq[FlavorSpurion]
y_{U} \sim ( \bar{\Yfund}, \Yfund, 1),
\quad
y_{D} \sim ( \bar{\Yfund}, 1, \Yfund).
\eeq
The observation of \Refs{TechniGIM} is that \FCNCs will be 
suppressed by a GIM mechanism if the only flavor violation 
above the compositeness scale comes in the form of two spurions with 
the transformation properties of $y_{U,D}$ above.
This is because all terms proportional to a single power of $y_{U,D}$
in the effective lagrangian at the weak scale can be simultaneously
diagonalized, so the only \FCNC effects are proportional to
combinations such as
$y_{U}^{\dagger} y\sub{U} y_{D}^{\dagger}$, which are not diagonal in 
the mass eigenstate basis.
However, such terms give rise to GIM-suppressed \FCNCs for the light 
quarks that are not in conflict with current bounds.

To see how this works, consider the example of $B \to X_{s}\ga$.%
\footnote{A related analysis in the context of extended 
technicolor theories is given in \Ref{RandallSundrum}.}
According to the spurion analysis outlined above, the leading 
contribution comes from operators in the effective theory below the
scale $\La$ such as
\beq
\de\scr{L}_{\rm eff} &\sim
\frac{g_{1}}{\La^{2}}
(y\sub{D} y_{U}^{\dagger} y\sub{U})^{\bar{k}}_{j}
\bar{d}\sub{R\bar{k}} \si^{\mu\nu} H\sub{D} Q_{L}^{j} B_{\mu\nu}
\nonumber\\
\eql{bsgammaOp}
&\sim \frac{g_{1}}{\La^{2}}\,
\frac{V_{ts} m_{t} m_{b}}{(v / \sqrt{2})^{2}}\,
\frac{v}{\sqrt{2}}\,
\bar{b}\sub{R} \si^{\mu\nu} s\sub{L} B_{\mu\nu} + \cdots
\eeq
Here $Q_{L}$ and $d_{R}$ are quark (fermion) fields, $H_{D}$ is a 
Higgs doublet (scalar) field, and $B_{\mu\nu}$ is the $\U1_{Y}$ 
field strength.
Operators proportional to
$y_{D}$ or $y\sub{D} y_{D}^{\dagger} y\sub{D}$ are flavor-diagonal, and do 
not contribute to \FCNCs.
The one-loop standard model contribution can be summarized by an 
operator of the form \Eq{bsgammaOp} in the effective theory below the 
weak scale, and the ratio of the coefficients is%
\footnote{For the standard model contribution and the definition of $A$,
see \Ref{InamiLim}.}
\beq
\frac{C\rsub{eff}}{C\rsub{std}}
\sim \frac{16 \pi^{2} v^{2}}{\La^{2}}\,
\frac{1}{A(m_{t}^{2} / M_{W}^{2})}
\sim6.0 \left( \frac{\La}{2\TeV} \right)^{-2}.
\eeq
The standard and non-standard contributions depend on the same 
combinations of quark masses and mixing angles because of the GIM 
mechanism.
We see that the non-standard contribution appears to be 
larger than the standard model contribution, but there are large
uncertainties in this estimate.

In the models we construct below, it turns out that the 
flavor spurions arise in the combination
$y_{U} H_{U}$ and $y_{D} H_{D}$, where $H_{U,D}$ are elementary 
Higgs fields.
We will refer to this as an ``enhanced'' GIM mechanism.
In these theories, the leading contribution to $B \to X_{s}\ga$ 
comes from
\beq
\de\scr{L} = \frac{g_{1}}{\La^{4}}
(y\sub{D} y_{U}^{\dagger} y\sub{U})^{\bar{k}}_{j}
\bar{d}_{R\bar{k}} \si^{\mu\nu} H\sub{D} H_{U}^{\dagger} H\sub{U}
Q_{L}^{j} B_{\mu\nu},
\eeq
which gives
\beq
\frac{C\rsub{eff}}{C\rsub{std}}
\sim \frac{16 \pi^{2} v^{2}}{\La^{2}}\,
\frac{(v / \sqrt{2})^{2}}{\La^{2}}\,
\frac{1}{A(m_{t}^{2} / M_{W}^{2})}
\sim 0.05 \left( \frac{\La}{2\TeV} \right)^{-4}.
\eeq

This pattern is repeated for other \FCNC observables.
The results are given in Table 1.
We conclude that the unenhanced GIM mechanism is marginally allowed,
while the enhanced GIM mechanism is completely safe.

\newcommand{\bigstrut}{\vphantom{\displaystyle\frac{1}{2}}}
\begin{table}
\begin{center}
\begin{tabular}{|c|c|c|}
\hline
           & unenhanced & enhanced \\
observable & GIM        & GIM \\
\hline\hline
$\bigstrut B \to X_{s}\ga$ &
$6 \left( \frac{\La}{2\TeV} \right)^{-2}$ &
$5 \times 10^{-2} \left( \frac{\La}{2\TeV} \right)^{-4}$ \\
\hline
$\bigstrut B_{s} \to X\mu^{+}\mu^{-}$ &
$0.25 \left( \frac{\La}{2\TeV} \right)^{-2}$ &
$2 \times 10^{-3} \left( \frac{\La}{2\TeV} \right)^{-4}$ \\
\hline
$\bigstrut B^{0}$--$\bar{B}^{0}$ mixing &
$4.5 \left( \frac{\La}{2\TeV} \right)^{-2}$ &
$3 \times 10^{-2} \left( \frac{\La}{2\TeV} \right)^{-4}$ \\
\hline
$\bigstrut K^{0}$--$\bar{K}^{0}$ mixing &
$2.5 \left( \frac{\La}{2\TeV} \right)^{-2}$ &
$2 \times 10^{-2} \left( \frac{\La}{2\TeV} \right)^{-4}$ \\
\hline
\end{tabular}
\capt{The ratio of coefficients $C_{\rm eff} / C_{\rm std}$
of dimension-6 \FCNC operators in the effective theory below the weak scale.}
\end{center}
\end{table}

A GIM mechanism such as the one described here can occur naturally in 
a composite model because the extra gauge symmetry of these models
generally leads to an enlarged accidental symmetry group.
These accidental symmetries may forbid flavor violation other than that 
parameterized by $y_{U,D}$.
Accidental flavor conservation is a feature of the standard model,
but not of the \MSSM.
We will also see that this feature arises in some of the models we 
construct in the next section.

\subsection{CP and B Violation}
\newcommand{\thang}{\ensuremath{\vartheta\rsub{QCD}}}
\newcommand{\ecm}{\ {\rm e}\,\,{\rm cm}}
In composite models, the extra accidental symmetries that occur may 
also suppress phenomenologically dangerous \CP and \B-violating 
processes \cite{EffSUSY}.
Constraints on \B violation are particularly severe:
even dimension-5 operators suppressed by $\sim 1/M\rsub{Planck}$ give
rise to unacceptably large \B violation.
It is therefore attractive if a model can forbid \B-violating 
operators up to dimension 5 by accidental symmetries.
In models where the quarks are composite, there is a simple mechanism 
for such accidental symmetries.
In the \MSSM, there are dimension-4 \B-violating operators are
superpotential of the form
$\bar{U}\bar{U}\bar{D}$ and $L Q \bar{D}$.
If the quarks are composites of preons $P$ with the quantum numbers
$P P$ (say), then the leading \B violating operators have the form
$\de W \sim (P P)^{3} / M^{3}$,
provided there are no other fields carrying baryon number.
This is more than sufficient to protect the theory from 
baryon number non-conservation arising at the Planck scale.

\Susc extensions of the standard model typically have extra 
\CP phases that can give dangerously large contributions to 
electric dipole moments \cite{CPSUGRA}.
Accidental symmetries can allow these phases to be rotated away.%
\footnote{The strong \CP problem can also be solved in this way
(the Nelson--Barr mechanism \cite{NelsonBarr}).}
Models with composite quarks and leptons generally have extra 
accidental symmetries, and therefore in general fewer \CP phases,
which may explain the smallness of observed \CP violation.

\subsection{Supersymmetry Breaking}
Since \susy is broken at a scale where the theory is strongly 
interacting, we cannot hope to get the \susy breaking dynamics fully 
under theoretical control.
However, with the recent advances in understanding non-perturbative 
effects in \susc gauge theories, we can usually determine with some
confidence whether or not \susy is broken in a particular model, even
if the model is strongly coupled.
This is something that must be addressed in each individual model.

\subsection{Light Gravitino}
Another interesting feature of this class of models is that the 
lightest \susc particle is the gravitino, with couplings suppressed by
$\sqrt{F} \sim 1 \TeV$.
As has been much discussed in the recent literature, this may explain 
the CDF $ee\ga\ga + \hbox{missing\ energy}$ event \cite{CDF}.
In this scenario, the events proceed via the decay of the 
next-lightest \susc particle (NLSP)
$N \to \chi\ga$, where $N$ is the NLSP and $\chi$ is the 
gravitino.
If the NLSP is the Bino (as suggested above), then the NLSP lifetime is
\cite{ExpGaG}
\beq
c\tau \sim 10^{-12}~{\rm m}\,
\left( \frac{m_{\twi{B}}}{100\GeV} \right)^{-5}
\left( \frac{\sqrt{F}}{1\TeV} \right)^{4},
\eeq
which is too small to be observed as a displaced vertex.
(In gauge-mediated \susy breaking models, there is also a light 
gravitino with $\sqrt{F} \sim 10$ to $100 \TeV$, and the NLSP decay can 
give rise to an observable displaced vertex \cite{ExpGaG}.)

\subsection{Summary}
Although not commonly considered in the literature,
models in which \susy and \ew symmetry are broken at 
the same scale are phenomenologically viable.
The main features of such models are:
$(i)$ quark compositeness at the scale $\La \sim \hbox{few\ TeV}$;
$(ii)$ \susy breaking at a scale $\sqrt{F} \sim \hbox{few\ TeV}$
with a light gravitino;
$(iii)$ a rich spectrum of strongly-interacting states at the TeV
scale.
It is difficult to make detailed quantitative predictions for the
superpartner spectrum and interactions in this class of models,
since they are strongly coupled.
Although these models have some features in common with technicolor 
theories, \FCNCs can be easily suppressed by an ``enhanced'' GIM
mechanism, or by approximate flavor symmetries.

One potentially serious problem for this class of models is precision 
\ew tests that show that experimental deviations from standard model
predictions are small.
This is often cited as evidence that the \ew symmetry breaking sector 
is weakly coupled.
In the present class of models, the compositeness scale $\La$ and 
the \susy breaking scale $\sqrt{F}$ are an order of magnitude larger
than the weak scale, which will reduce the corrections.
This issue deserves further study, but we will not address it here.

\section{A Model with Three Composite Generations}
We now present an explicit model with three composite generations of
quarks and leptons, which illustrates many of the ideas described above.
The model is based on the charge assignments of \Ref{us}.
This model has an unrealistic fermion mass spectrum, and eliminating 
unwanted light states requires the addition of many new interactions.
However, simpler versions of this model exhibit interesting features
such as an enhanced GIM mechanism and near-universal masses for all 
up- and down-type squarks.

\subsection{Field Content}
We describe the model by giving the representations of the fields under
the group
\beq[theGroup]
\bal
\SU{8}_{H} &\times \SU{3}_{C} \times \SU{2}_{W} \times \U1_{Y}
\\
&\times \bigl[ \SU{3}_{Q} \times \SU{3}_{U} \times \SU{3}_{D}
\times \U1_{B} \times \U1_{L} \bigr],
\eal\eeq
where $\SU{8}_{H}$ is a strong ``hypercolor'' gauge group,
$\SU{3}_{C} \times \SU{2}_{W} \times \U1_{Y}$ is the usual standard 
model gauge group, and the groups in brackets are global symmetries,
some of which will be explicitly broken.
We explicitly give the baryon and lepton number assignments to 
show that these are anomaly-free conserved quantum numbers in this
model.
The hypercolored sector of the theory consists of the ``preon'' fields
\beq\bal
P_{Q} &\sim (\Yfund, 1, \Yfund)_{0}
\times (\Yfund, 1, 1)_{\frac{1}{8},\, \frac{1}{8}}
\\
\bar{P}_{U} &\sim (\bar{\Yfund}, 1, 1)_{-1}
\times (1, \bar{\Yfund}, 1)_{-\frac{1}{8},\, -\frac{1}{8}}
\\
\bar{P}_{D} &\sim (\bar{\Yfund}, 1, 1)_{1}
\times (1, 1, \bar{\Yfund})_{-\frac{1}{8},\, -\frac{1}{8}}
\\
P_{C} &\sim (\Yfund, \bar{\Yfund}, 1)_{-\frac{1}{3}}
\times (1, 1, 1)_{-\frac{5}{24},\, \frac{1}{8}}
\\
\bar{P}_{C} &\sim (\bar{\Yfund}, \Yfund, 1)_{\frac{1}{3}}
\times (1, 1, 1)_{\frac{5}{24},\, -\frac{1}{8}}
\eal\eeq
The $\SU{8}_{H}$ interactions become strong at a scale $\La \sim 1\TeV$.
To make the model realistic, additional fields and interactions will
be required, but these will be weakly coupled at the scale $\La$.
The $\SU{8}_{H}$ group has the right number of matter fields to 
confine smoothly without breaking chiral symmetries \cite{Seiberg}.
This means that the effective theory below the scale $\La$ consists
of composite fields with a dynamically generated \spot.
The composite ``meson'' fields are
\beq\bal
Q &\sim \bar{P}_{C} P_{Q} \sim (1, \Yfund, \Yfund)_{\frac{1}{3}}
\times (\Yfund, 1, 1)_{\frac{1}{3},\, 0}
\\
\bar{U} &\sim \bar{P}_{U} P_{C} \sim (1, \bar{\Yfund}, 1)_{-\frac{4}{3}}
\times (1, \bar{\Yfund}, 1)_{-\frac{1}{3},\, 0}
\\
\bar{D} &\sim \bar{P}_{D} P_{C} \sim (1, \bar{\Yfund}, 1)_{\frac{2}{3}}
\times (1, 1, \bar{\Yfund})_{-\frac{1}{3},\, 0}
\\
\bar{\Phi}_{U} &\sim \bar{P}_{U} P_{Q} \sim (1, 1, \Yfund)_{-1}
\times (\Yfund, \bar{\Yfund}, 1)_{0,\, 0}
\\
\bar{\Phi}_{D} &\sim \bar{P}_{D} P_{Q} \sim (1, 1, \Yfund)_{1}
\times (\Yfund, 1, \bar{\Yfund})_{0,\, 0}
\\
A &\sim (\bar{P}_{C} P_{C})_{8} \sim (1, \hbox{\bf Ad}, 1)_{0}
\times (1, 1, 1)_{0,\, 0}
\\
T &\sim (\bar{P}_{C} P_{C})_{1} \sim (1, 1, 1)_{0}
\times (1, 1, 1)_{0,\, 0}
\eal\eeq
where {\bf Ad} denotes the adjoint representation.
The composite ``baryon'' fields are
\beq\bal
L &\sim P_{C}^{3} P_{Q}^{5} \sim (1, 1, \Yfund)_{-1}
\times (\bar{\Yfund}, 1, 1)_{0,\, 1}
\\
\bar{E} &\sim \bar{P}_{C}^{3} \bar{P}_{U}^{2} \bar{P}_{D}^{3}
\sim (1, 1, 1)_{2}
\times (1, \Yfund, 1)_{0,\, -1}
\\
\bar{N} &\sim \bar{P}_{C}^{3} \bar{P}_{U}^{3} \bar{P}_{D}^{2}
\sim (1, 1, 1)_{0}
\times (1, 1, \Yfund)_{0,\, -1}
\\
X &\sim P_{C}^{2} P_{Q}^{6}
\sim (1, \Yfund, 1)_{-\frac{2}{3}}
\times (1, 1, 1)_{\frac{1}{3},\, 1}
\\
\bar{X} &\sim \bar{P}_{C}^{2} \bar{P}_{U}^{3} \bar{P}_{D}^{3}
\sim (1, \bar{\Yfund}, 1)_{\frac{2}{3}}
\times (1, 1, 1)_{-\frac{1}{3},\, -1}
\eal\eeq
The dynamical \spot is
\beq[Wdyn]
\bal
W_{\rm dyn} &=
L \bar{\Phi}_{U} \bar{E} 
+ L \bar{\Phi}_{D} \bar{N}
+ L Q \bar{X}
+ X \bar{U} \bar{E}
+ X \bar{D} \bar{N}
+ X A \bar{X} + X T \bar{X}
\\
& \quad - \hbox{determinant},
\eal\eeq
where ``determinant'' denotes terms proportional to 9 powers of the 
composite meson fields whose precise form will not play an important
role in our discussion.
We see that the composite fields include three full generations of
quarks and leptons ($Q$, $\bar{U}$, $\bar{D}$, $L$, $\bar{E}$, plus
right-handed neutrinos $\bar{N}$) and several additional fields:
$\bar{\Phi}_{U,D}$ are ``flavored'' Higgs fields,
$A$ is a color octet,
$T$ is a singlet,
and $X, \bar{X}$ are leptoquarks.

This is a promising starting point for constructing a model of the 
type described in the previous section.
However, the model as described so far is clearly far from realistic.
For example, \susy is unbroken, and the model has a moduli space
of \susc vacua.
Also, there is an exact flavor symmetry that forbids quark and lepton
masses.

\subsection{Supersymmetry and Electroweak Symmetry Breaking}
To break \susy and \ew symmetry
in this model, we add the following terms to the 
\spot of the model above the scale $\La$:
\beq[ORSector]
\de W = \sum_{j = 1}^{3} \la_{j} S_{j} \left(
P_{C} \bar{P}_{C} + h_{j} H_{U} H_{D} + \ka_{j}
\right),
\eeq
where we have introduced new elementary fields
\beq\bal
S_{j} &\sim (1, 1, 1)_{0} \times (1, 1, 1)_{0,\, 0},
\\
H_{U} &\sim (1, 1, \Yfund)_{1} \times (1, 1, 1)_{0,\, 0},
\\
H_{D} &\sim (1, 1, \Yfund)_{-1} \times (1, 1, 1)_{0,\, 0}.
\eal\eeq
That is, $S_{j}$ are singlets,
and $H_{U,D}$ are elementary Higgs fields.
This corresponds to an \OR sector that breaks \susy and \ew symmetry
even in the limit where the strong $\SU{8}_{H}$ gauge interactions
are turned off.
This is somewhat unattractive, since the scale of \susy breaking is 
essentially put in by hand, in the form of the dimensionful
parameters $\ka_{j}$.
We note however that having the $\ka_{j}$ small is natural,
since the theory has additional \U1 symmetries in the limit
$\ka_{j} \to 0$.
Also, both the \susy and \ew symmetry
breaking scales are determined by the same parameters $\ka_{j}$.
What is not explained is why $\ka_{j} \sim \La^{2}$.
An attractive alternative would be to consider models such as those 
in \Refs{ISSIT} that dynamically break \susy at the compositeness scale,
and introduce new interactions that break \ew symmetry at one loop
at a scale $\La / (4\pi) \sim M_{W}$.

\Eq{ORSector} also preserves a $\U1_{R}$ symmetry under which the 
fields $S_{j}$ have charge $+2$ and all other fields are uncharged.
When $\avg{S} \ne 0$ (see below) this will give rise to an $R$-axion 
unless the $\U1_{R}$ symmetry is explicitly broken.%
\footnote{It may be necessary to break $R$ symmetries explicitly to
solve the cosmological constant problem \cite{BPR}.}
For example, the $\U1_{R}$ symmetry may be broken by higher-dimension
terms in the \Kahler potential.

In the presence of the strong $\SU{8}_{H}$ gauge interactions,
the terms in \Eq{ORSector} give rise to the effective \spot below
the compositeness scale
\beq[effOR]
\de W_{\rm eff} = \sum_{j = 1}^{3} \la_{j} S_{j} \left(
\La T + h_{j} H_{U} H_{D} + \ka_{j}
\right),
\eeq
and it is easily checked that
\beq
\avg{T} \sim \frac{\ka}{\La},
\qquad
\avg{H_{U,D}} \sim \ka^{1/2},
\qquad
\avg{\left. S_{j} \right|_{\th\th}} \sim \ka.
\eeq
This effective theory is valid only for $\avg{T} \lsim \La$, but the 
fact that the theory defined by \Eq{ORSector} breaks \susy at tree level
excludes the possibility that there is a \susc vacuum for large values
of $\avg{T}$.

The \vev $\avg{T} \ne 0$ also gives rise to masses for 
the leptoquarks through the term $X T \bar{X}$ in the dynamical 
\spot \Eq{Wdyn}.
Although the leptoquark masses are at most of order a TeV, the
leptoquark masses preserve the full flavor symmetry, since they arise from
flavor-blind strong dynamics.
There are therefore no problems with \FCNCs from leptoquark exchange.

\subsection{Superpartner Masses}
We now show that this model generates terms in the effective \Kahler 
potential analogous to \Eqs{SquarkKeff} and \eq{GauginoKeff} that 
give rise to superpartner masses.
The combination of fields
\beq
S \equiv \la_{j} S_{j}
\eeq
acts as a field-dependent mass term for the preons
$P_{C}, \bar{P}_{C}$, and we can 
use standard spurion analysis to work out the dependence on $S$ 
in the effective theory below the scale $\La$.
We use dimensional analysis to estimate the size of the various terms
in the effective theory, but we 
must be careful to include all small suppression factors.
In the limit where all the standard model gauge couplings are turned off,
the $\SU{8}_{H}$ gauge theory has a large anomaly-free global symmetry
$\SU{9} \times \SU{9} \times \U1 \times \U1_{R}$.
To keep track of this symmetry, we write the composite fields as
\beq
M^{J}{}_{\bar{K}} = P^{J} \bar{P}_{\bar{K}},
\qquad
B_{J} = (P \cdots P)_{J},
\qquad
\bar{B}^{\bar{K}} = (\bar{P} \cdots \bar{P})^{\bar{K}},
\eeq
where $J, \bar{K} = 1, \ldots, 9$ run over the various preon
fields.
The coupling of $S$ to the preons in the \spot can be written
\beq
S P_{C} \bar{P}_{C} = m^{\bar{K}}{}_{J} P^{J} \bar{P}_{\bar{K}}.
\eeq

With this notation, we can write terms in the effective \Kahler
potential that are manifestly invariant under the full global 
symmetries of the strong interactions.
The terms
\beq[Ksquark]
\bal
\de K_{\rm eff} &= \frac{1}{\La^{2}} \tr(m^{\dagger} m)
\tr(M^{\dagger} M)
\\
&= \frac{1}{\La^{2}} S^{\dagger} S
\left( Q^{\dagger} Q + \bar{U}^{\dagger} \bar{U} + \cdots \right)
\eal\eeq
and
\beq\bal
\de K_{\rm eff} &= \frac{1}{\La^{2}} \tr(m^{\dagger} m)
\left( B^{\dagger} B + \bar{B}^{\dagger} \bar{B} \right)
\\
&= \frac{1}{\La^{2}} S^{\dagger} S
\left( L^{\dagger} L + \bar{E}^{\dagger} \bar{E} + \cdots \right)
\eal\eeq
give rise to squark and slepton masses provided
$\avg{\left. S\right|_{\th\th}} \ne 0$.
Notice that \Eq{Ksquark} gives rise to the same mass for all up- and 
down-type squarks;
squark splittings come from weak gauge interactions and small flavor 
violations, so all squarks are nearly degenerate in this model.
The term
\beq\bal
\de K_{\rm eff} &= \frac{g^{2}}{\La^{5}}
\tr(m^{\dagger} m) \tr(m^{\dagger} M^{\dagger}) \tr(WW)
\\
&= \frac{g^{2}}{\La^{5}} S^{\dagger 2} S T^{\dagger} \tr(WW) + \cdots
\eal\eeq
is allowed by all symmetries (including $\U1_{R}$ symmetries)
and gives rise to gaugino masses provided that
$\avg{S}, \avg{T} \ne 0$ as
well as $\avg{\left. S\right|_{\th\th}} \ne 0$.

\subsection{Flavor Structure}
Even without additional flavor violation, this model gives rise to 
flavor-changing neutral currents from operators such as \cite{us}
\beq\bal
\de K\rsub{eff} &= \frac{1}{\La^{2}} \left[
B M M^{\dagger} B^{\dagger}
+ \bar{B}^{\dagger} M^{\dagger} M \bar{B} \right]
\\
&= \frac{1}{\La^{2}} \left[
L\sub{j} Q^{j} Q^{\dagger}_{k} L^{\dagger k}
+ \bar{E}^{\dagger}_{\bar{k}} \bar{U}^{\dagger\bar{k}}
\bar{U}\sub{\bar{\ell}} \bar{E}^{\bar{\ell}} + \cdots \right],
\eal\eeq
where we have shown the flavor indices for clarity.
The most stringent bound on such operators comes from
$D^{0} \to e^{-} \mu^{+}$, and gives a bound $\La \lsim 4\TeV$ 
\cite{us}.

To give masses to the quarks and leptons, we must break the flavor 
symmetry $\SU{3}_{Q} \times \SU{3}_{U} \times \SU{3}_{D}$.
The only dimension-4 terms compatible with the gauge interactions that 
can break the flavor symmetries are
\beq[FlavorTerms]
\de W_{0} = y_{U} H_{U} P_{Q} \bar{P}_{U}
+ y_{D} H_{D} P_{Q} \bar{P}_{D},
\eeq
where $y_{U,D}$ are dimensionless couplings that transform under the
flavor symmetry as in \Eq{FlavorSpurion}.
We see that this model as described so far has a natural enhanced GIM 
mechanism enforced by accidental symmetries.

In the notation of the previous subsection, the terms in
\Eq{FlavorTerms} correspond to additional field-dependent mass terms 
for the preon fields.
We therefore have terms in the effective \Kahler potential such as
\beq\bal
\de K_{\rm eff} &= \frac{1}{\La^{4}}
\tr(M^{\dagger} m^{\dagger} M^{\dagger} M m M)
\\
&= \frac{1}{\La^{4}}
T^{\dagger 2} S^{\dagger}
\left( y_{U} Q H_{U} \bar{U} + y_{D} Q H_{D} \bar{D} \right)
+ \cdots
\eal\eeq
and
\beq\bal
\de K_{\rm eff} &= \frac{1}{\La^{3}}
( B m^{\dagger} \bar{B} ) \tr(m^{\dagger} m)
\\
&= \frac{1}{\La^{3}}
S^{\dagger} S \left( y_{U}^{\dagger} L H_{U}^{\dagger} \bar{E}
+ y_{D}^{\dagger} L H_{D}^{\dagger} \bar{N} \right) + \cdots
\eal\eeq
These give rise to quark and lepton masses provided that
$\avg{S}, \avg{T} \ne 0$ and
$\avg{\left. S\right|_{\th\th}} \ne 0$.

This model predicts that the charged lepton masses are 
proportional to the up-type quark masses.
This is a direct consequence of the fact that the composite quark and 
lepton fields transform under the same flavor symmetry.
Although this is definitely not realistic,
we will continue to analyze this model to see what we can learn from 
it.

The flavor structure of this model leads to another phenomenological
problem:
the fields $\bar{\Phi}_{U,D}$ also transform under the flavor group, and 
the masses of the fermion components of these superfields (``higgsinos'') 
are protected by the flavor symmetries.
This means that there will be additional charged fermions with masses 
of order the light quark and lepton masses, which is clearly excluded.

In order to
avoid the existence of the light higgsinos, we must add more flavor 
structure to the model.
This will destroy the GIM mechanism, but we will see that the model can
be made realistic without fine-tuning by assuming the existence of
approximate flavor symmetries.
A similar idea was discussed in the context of 
weakly-coupled multi-Higgs models in
\Ref{FlavorScalars}.
In the present model, the flavor symmetries act on the Higgs fields as well
as the quark fields, since there is only one flavor group at the preon
level.

We therefore introduce additional elementary flavored Higgs fields
\beq\bal
\Phi_{U} &\sim (1, 1, \Yfund)_{1}
\times (\bar{\Yfund}, \Yfund, 1)_{0,\, 0}
\\
\Phi_{D} &\sim (1, 1, \Yfund)_{-1}
\times (\bar{\Yfund}, 1, \Yfund)_{0,\, 0}
\eal\eeq
We then add the couplings to the lagrangian above the compositeness 
scale%
\footnote{The symmetries also allow flavor-violating terms of the form
$H_{U} \Phi_{D}$ and $H_{D} \Phi_{U}$.
Including such terms does not affect the results below.}
\beq[PhiMass]
\de W_{0} = ( \la_{U} )^{\bar{k} m}_{j \bar{\ell}}
P_{Q}^{j} \bar{P}_{U \bar{k}} ( \Phi_{U} )^{\bar{\ell}}_{m}
+ ( \la_{D} )^{\bar{k} m}_{j \bar{\ell}}
P_{Q}^{j} \bar{P}_{D \bar{k}} ( \Phi_{D} )^{\bar{\ell}}_{m},
\eeq
where we show the flavor indices explicitly.

Below the compositeness scale, these couplings give masses to the 
fields $\Phi_{U,D}$ and $\bar{\Phi}_{U,D}$%
\footnote{With the addition of these fields, there are 44 weak 
doublets, and the $\SU{2}_{W}$ gauge coupling constant diverges at a
scale $\sim 130 \La$.}
\beq
\de W_{\rm eff} = \La ( \la_{U} )^{\bar{k} m}_{j \bar{\ell}}
(\bar{\Phi}_{U})^{j}_{\bar{k}} ( \Phi_{U} )^{\bar{\ell}}_{m}
+ \La ( \la_{D} )^{\bar{k} m}_{j \bar{\ell}}
(\bar{\Phi}_{D})^{j}_{\bar{k}} ( \Phi_{D} )^{\bar{\ell}}_{m}.
\eeq
If the couplings $\la_{U,D}$ are chosen arbitrarily, this model will 
certainly have \FCNCs.
However, arbitrary values for the couplings also do not explain the 
observed pattern of quark and lepton masses.
We therefore follow \Ref{FlavorScalars} and assume
that the  smallness of the first two generations of quark and lepton
masses is due to approximate \U1\ flavor symmetries.
We therefore write
\beq
( \la_{U} )^{\bar{k} m}_{j \bar{\ell}} = \la_{U0} 
\de^{\bar{k}}_{\bar{\ell}} \de^{m}_{j} + O(\ep\sub{Qj} \ep\sub{U\bar{k}} 
\ep\sub{U\bar{\ell}} \ep\sub{Qm}),
\eeq
\etc, where the $\ep$'s are suppression factors associated with the 
various generations.
The first term is invariant under the flavor symmetries, so it is 
natural that $\la_{U0} \sim 1$, so the fields $\Phi_{U,D}$ so 
$\bar{\Phi}_{U,D}$ get masses of order $\La$.

The new interaction terms give rise to \FCNC operators at the weak
scale such as
\beq
\de \scr{L}_{\rm eff} &= \frac{1}{\La^{4}}
( \la_{D} \Phi_{D} )^{\bar{k}}_{j}
( \la_{D} \Phi_{D} )^{\dagger\ell}_{\bar{m}}
( \bar{Q}\sub{L\ell} \ga^{\mu} Q_{L}^{j} )
( \bar{d}_{R\bar{k}} \ga_{\mu} d_{R}^{\bar{m}} )
\nonumber\\
&= \frac{\la_{D0}^{2}}{\La^{4}}
( \Phi\sub{D} )^{1}_{2}
( \Phi_{D}^{\dagger} )^{1}_{2}
( \bar{d}_{L} \ga^{\mu} s_{L} )
( \bar{d}_{R} \ga_{\mu} s_{R} )
+ \cdots
\eeq
If $\avg{\Phi_{D}} \sim v$ this will give rise to unacceptably 
large $K^{0}$--$\bar{K}^{0}$ mixing.
However, $\avg{\Phi_{D}}$ can be naturally zero because of 
the accidental $\U1_{R}$ symmetry under which $\Phi_{D}$ has 
$R = +2$ and all other fields are uncharged.
This $\U1_{R}$ may be unbroken, and leads to no unacceptable 
phenomenological consequences.

There are also contributions to \FCNCs due to
the exchange of heavy $\Phi_{U,D}$ and $\bar{\Phi}_{U,D}$ scalars.
To work out the couplings of the fields $\Phi_{U,D}$ to quarks,
we note that \Eq{PhiMass} has the form of an additional field-dependent
contribution to the preon mass.
Below the compositeness scale, this will give rise to terms such as
\beq
\de K_{\rm eff} &= \frac{1}{\La^{4}} \tr(m^{\dagger} M^{\dagger})
\tr(m M M^{\dagger} M) + \hc
\nonumber\\
&= \frac{1}{\La^{4}} S^{\dagger} T^{\dagger} \left[
(\la_{U} \Phi_{U})^{\bar{k}}_{j} Q^{j} T^{\dagger} 
\bar{U}_{\bar{k}}
+ (\la_{D} \Phi_{D})^{\bar{k}}_{j} Q^{j} T^{\dagger} 
\bar{D}_{\bar{k}} \right] + \cdots,
\eeq
which gives rise to Yukawa couplings between the fields 
$\Phi_{U,D}$ and the quarks if
$\avg{T} \ne 0$, $\avg{\left.S\right|_{\th\th}} \ne 0$.
In this case, the effective lagrangian at the weak scale includes
\beq
\de\scr{L}_{\rm eff} = (\la_{U} \Phi_{U})^{\bar{k}}_{j}
\bar{u}_{R\bar{k}} Q_{Lj} +
(\la_{D} \Phi_{D})^{\bar{k}}_{j}
\bar{d}_{R\bar{k}} Q_{Lj} + \hc
\eeq
The exchange of $\Phi_{U,D}$ and $\bar{\Phi}_{U,D}$ scalars induces 
operators below the weak scale such as
\beq
\de\scr{L}_{\rm eff} &\sim \frac{1}{m_{\Phi}^{2}}
( \la_{D} )^{\bar{k} n}_{j \bar{p}}
( \la_{D}^{\dagger} )^{\bar{p} \ell}_{n \bar{m}}
( \bar{d}\sub{R\bar{k}} Q_{L}^{j} )
( \bar{Q}\sub{L\ell} d_{R}^{\bar{m}} )
\nonumber\\
&\sim \frac{\la_{D0}}{m_{\Phi}^{2}} \Bigl[
\ep\sub{D1} \ep\sub{Q2} \ep\sub{Q1} \ep\sub{D2}
( \bar{d}\sub{R} s\sub{L} )
( \bar{d}\sub{L} s\sub{R} )
+ \ep\sub{Q1} \ep\sub{Q3} \ep\sub{D1} \ep\sub{D3}
(\bar{d}\sub{R} b\sub{L})
(\bar{d}\sub{L} b\sub{R}) \Bigr].
\eeq
Using the suppression factors of \Ref{FlavorScalarstoo}, this
gives rise to contributions to $K^{0}$--$\bar{K}^{0}$ and
$B^{0}$--$\bar{B}^{0}$ mixing of order
\beq\bal
\hbox{$K^{0}$--$\bar{K}^{0}$:} &\quad
\frac{C\rsub{eff}}{C\rsub{std}}
\sim 0.8 \left( \frac{m_{\Phi}}{500\GeV} \right)^{-2},
\\
\hbox{$B^{0}$--$\bar{B}^{0}$:} &\quad
\frac{C\rsub{eff}}{C\rsub{std}}
\sim 1 \left( \frac{m_{\Phi}}{500\GeV} \right)^{-2}.
\eal\eeq
This means that the scalars can contribute to $K^{0}$--$\bar{K}^{0}$ and
$B^{0}$--$\bar{B}^{0}$ mixing at the same level as the standard model
box diagrams.
As discussed in \Ref{FlavorScalarstoo}, this implies that the 
scalars can make large contributions to observed \CP violation, and 
suggests that \CP may be an approximate symmetry at the weak scale.

As it stands, the neutrinos have Dirac masses of order the 
corresponding down-type quark masses, which is clearly unacceptable.
To avoid this, can make the right-handed neutrino $\bar{N}$ 
heavy.
We therefore assume that there are extra particles with masses
$M \gsim \La$ that give rise to interactions at the compositeness 
scale of the form
\beq[NewNu]
\de W = \frac{1}{M^{4}}
N \bar{P}_{C}^{3} \bar{P}_{U}^{2} \bar{P}_{D},
\eeq
where $N$ is an elementary neutrino field
\beq
N \sim (1, 1, 1)_{0} \times (1, 1, \bar{\Yfund})_{0,\, 1}.
\eeq
(Note that lepton number is not violated.)
Below the compositeness scale, this gives rise to the interaction
\beq
\de W_{\rm eff} = m_{N} N \bar{N},
\eeq
where $m_{N} \sim \La^{5} / M^{4}$.
The left-handed neutrinos are exactly massless in this model, but 
their couplings to weak currents are suppressed by a factor
$1 - \ep_{j}$, with $\ep_{j} \sim m_{dj}/ m_{N}$.
The most stringent bound on $m_{N}$ comes from comparing $\mu$ and 
$\tau$ weak decays, and gives $m_{N} \gsim 750\GeV$.
This means that the scale $M$ appearing in \Eq{NewNu} is close to $\La$, 
and we should include the states with mass $M$ in the dynamics at the 
scale $\La$.
There is no reason to expect this to change the dynamics 
qualitatively, and we will not pursue this further.

\subsection{Global Symmetries}
The leading contribution to \B violation in this model comes from 
dimension 7 operators, as discussed in Section 2, so this model 
gives an explanation for the absence of observed \B violation.
On the other hand, this model allows non-removable \CP phases in
interaction terms such as \Eq{ORSector} that can lead to observable 
electric dipole moments.
Therefore, this model does not explain the smallness of observed \CP 
violation.
However, as discussed above this model can be natural and viable
if \CP is an approximate symmetry.

\subsection{Summary}
This is not a pretty model.
In particular, it contains several dimensionful couplings
whose origin is not explained within the model.
On the other hand, the model does exhibit the feature that \susy and 
\ew symmetry breaking have the same dynamical origin.
Also, less realistic versions of this model are quite 
simple, and hold out the hope that better models exist.
The main point of constructing this model is to exhibit a specific
model that has compositeness and \susy breaking near the weak scale,
and avoids the existence of light superpartners and
flavor-changing neutral currents.
We hope that more elegant models can be constructed that address the
unattractive features of the model considered above.

\section{A Model with Composite Right-handed Down Quarks}
Much of the complication of the model of the previous section arose 
from the need to introduce additional flavor violation in order to 
make unwanted flavored composite fermions heavy.
In this section, we construct a model that has extra flavored 
fermions that can be made heavy without introducing extra flavor
violation.
In this model, \FCNCs are suppressed by the enhanced GIM mechanism
described in Section 2.

Only the right-handed down quarks are composite in this model.
We give the representations of the fields under the group
\beq[theSmallerGroup]
\bal
\SU{3}_{H} &\times \SU{3}_{C} \times \SU{2}_{W} \times \U1_{Y}
\\
&\times \bigl[ \SU{3}_{D} \times \U1_{B} \bigr],
\eal\eeq
in a notation similar to the previous section.
The preon fields are
\beq\bal
\bar{P}_{D} &\sim (\bar{\Yfund}, 1, 1)_{\frac{2}{3}}
\times \bar{\Yfund}_{-\frac{1}{3}},
\\
\bar{P}_{1} &\sim (\bar{\Yfund}, 1, 1)_{-\frac{2}{3}}
\times 1_{1},
\\
P_{C} &\sim (\Yfund, \bar{\Yfund}, 1)_{0} \times 1_{0},
\\
P_{1} &\sim (\Yfund, 1, 1)_{0} \times 1_{0}.
\eal\eeq
As before, this theory confines smoothly at a scale $\La \sim 1\TeV$.
Below the scale $\La$, the light composite fields are the ``mesons''
\beq\bal
\bar{D} &\sim \bar{P}_{D} P_{C} \sim (1, \bar{\Yfund}, 1)_{\frac{2}{3}}
\times \bar{\Yfund}_{-\frac{1}{3}},
\\
\De &\sim \bar{P}_{D} P_{1} \sim (1, 1, 1)_{\frac{2}{3}}
\times \Yfund_{-\frac{1}{3}},
\\
X &\sim \bar{P}_{1} P_{C} \sim (1, \bar{\Yfund}, 1)_{-2}
\times 1_{1},
\\
T &\sim \bar{P}_{1} P_{1} \sim (1, 1, 1)_{-2}
\times 1_{1},
\eal\eeq
and ``baryons''
\beq\bal
\bar{B}_{D} &\sim \bar{P}_{D}^{2} P_{1} \sim
(1, 1, 1)_{-\frac{2}{3}} \times \Yfund_{\frac{1}{3}},
\\
\bar{B}_{1} &\sim \bar{P}_{D}^{3} \sim
(1, 1, 1)_{2} \times 1_{-1},
\\
B_{C} &\sim P_{C}^{2} P_{1} \sim
(1, \Yfund, 1)_{0} \times 1_{0},
\\
B_{1} &\sim P_{C}^{3} \sim
(1, 1, 1)_{0} \times 1_{0}.
\eal\eeq
This model has a dynamical superpotential
\beq[Wdyntoo]
W\rsub{dyn} = \bar{B}_{D} \bar{D} B_{C} + \bar{B}_{D} \De B_{1}
+ \bar{B}_{1} X B_{C} + \bar{B}_{1} T B_{1}
- \hbox{determinant},
\eeq
where ``determinant'' denotes terms proportional to 4 powers of the 
composite meson fields.

We see that this theory has unwanted flavored states $\De$ and 
$\bar{B}_{D}$.
However, unlike the model of the previous section, we do not have to 
introduce additional flavor violation to make the fermion components 
of these fields heavy.
From \Eq{Wdyntoo}, we see that if
$\avg{B_{1}} \ne 0$, then $\De$ and $\bar{B}_{D}$ will get a 
flavor-invariant Dirac mass.

One way to obtain $\avg{B_{1}} \ne 0$ is as follows.
Above the compositeness scale, we add the interactions
\beq
\de W_{0} = \frac{1}{M} S P_{C}^{3} + \ka S,
\eeq
where
\beq
S \sim (1, 1, 1)_{0} \times 1_{0}
\eeq
is an elementary singlet field.
The higher-dimension interaction above can be generated by integrating 
out unflavored states with mass $M$.
Below the compositeness scale, this gives rise to the interactions
\beq
\de W\rsub{eff} = \frac{\La^{2}}{M} S B_{1} + \ka S,
\eeq
which give
\beq
\avg{B_{1}} \sim \frac{\ka M}{\La^{2}}.
\eeq

We can break the $\SU{3}_{D}$ flavor symmetry by adding interactions 
at the compositeness scale such as
\beq[newWzero]
\de W_{0} = \frac{(y_{D})^{\bar{k}}_{j}}{M} Q^{j} H_{D}
(\bar{P}_{D \bar{k}} P_{C}),
\eeq
where $Q$ is an elementary quark field and $H_{D}$ is an elementary 
Higgs field.
Note that the flavor-breaking spurions appear together with
$H_{D}$, so this model has an enhanced GIM mechanism.
In order for the GIM mechanism to be natural at the scale $M$,
it is important that these interactions can
arise from integrating out unflavored states.
For example, we can introduce
\beq\bal
P_{Q} &\sim (\Yfund, \bar{\Yfund}, \Yfund)_{-1}
\times 1_{0},
\\
\bar{P}_{Q} &\sim (\bar{\Yfund}, \Yfund, \Yfund)_{1}
\times 1_{0},
\eal\eeq
and replace \Eq{newWzero} by
\beq
\de W_{0} = (y_{D})^{\bar{k}}_{j} P_{Q} \bar{P}_{D \bar{k}} Q^{j}
+ M P_{Q} \bar{P}_{Q}
+ \bar{P}_{Q} P_{C} H_{D}.
\eeq
Below the scale $M$, we can integrate out the fields $P_{Q}$ and 
$\bar{P}_{Q}$ to obtain \Eq{newWzero}.

Below the compositeness scale, \Eq{newWzero} gives rise to 
interactions
\beq
W\rsub{eff} = \frac{\La (y_{D})^{\bar{k}}_{j}}{M}
Q^{j} H_{D} \bar{D}_{\bar{k}},
\eeq
which generate down-quark masses and mixings.
The up-type quarks and leptons get masses from elementary Yukawa 
couplings.
$\La \ll M$ may explain why $m_{b} \ll m_{t}$ in this 
model.

We will not analyze this model in any more detail.
The point of presenting it is to illustrate that the 
extra flavored states that generically occur in composite models do 
not necessarily imply the existence of extra flavor violation that 
destroys the GIM mechanism.

\section{Conclusions}
We have described the general features of models in which \susy is 
broken near the weak scale by fields that carry \ew quantum numbers.
We have argued that in these theories some or all of the quarks and 
leptons are composite at a few TeV and scalar superpartner masses 
arise from non-perturbative effects.
These models predict that searches for compositeness (in the 
form of high-$p_{T}$ jet enhancement, for example) will yield positive
results at energies near the current experimental limits.
The lightest \susc particle in these theories is the gravitino
with stronger couplings than in gauge-mediated models, so that the 
signals are missing energy events rather than displaced vertices.
These models also predict a rich spectrum of strongly-interacting 
states at a few TeV, as in technicolor theories.

On the model-building side, we have seen that \FCNCs can be adequately
suppressed by a GIM mechanism or by approximate flavor symmetries.
While we have not constructed a completely satisfactory model, we 
believe that this framework represents an interesting new possibility 
for realizing \susy in nature.

The combination of \susy and strong dynamics has been considered 
previously by several authors.
Models that break \susy dynamically through strong interactions were 
considered already in the 1980's \cite{GaugeMediated}.
These models use a gauge-mediated mechanism for communicating \susy 
breaking to standard-model superpartners.
More recently, the framework of ``bosonic technicolor'' 
combined technicolor dynamics with \susy \cite{BosonicTC}.
In these models the \susy breaking takes place above the scale where
technicolor is strongly coupled.
Models with composite quarks and leptons were considered in
\Refs{NelsonStrassler} to explain flavor physics, and in
\Ref{EffSUSY} to explain \susy breaking.
The scenario advocated in \Ref{EffSUSY} differs from ours in several
respects:
$(i)$ the third generation is fundamentally different in their framework;
$(ii)$ \FCNCs in the first two generations are suppressed by a
combination of large quark masses and small couplings;
and
$(iii)$ the compositeness and \susy breaking scales are larger than
in our models.

\section{Acknowledgments}
I would like to thank R.N.~Mohapatra, M.~Schmaltz, and J.~Terning for
discussions.


\end{document}